\input harvmac

\lref\hhk{ J. Harvey, P. Horava, P. Kraus, ``D-Sphalerons 
 and the topology of string configuration space,''  
JHEP 0003 (2000) 021,  hep-th/0001143  } 
\lref\reyman{ G.Mandal and S.J.Rey
``A Note on D-Branes of Odd Codimensions from Noncommutative
Tachyons,'' hep-th/0008214 } 
\lref\hk{J. Harvey and P. Kraus ``Tensionless branes'' } 
\lref\klu{ J. Kluson ``D-Branes from N non-BPS D0-Branes,''hep-th/0009189. } 
\lref\hklm{ J.Harvey, P. Kraus, F. Larsen, E. Martinec, 
 ``D-branes and Strings as 
Non-commutative Solitons,'' hep-th/0005031,
JHEP 0007 (2000) 042 } 
\lref\senel{ A. Sen ``Fundamental Strings in Open
 String Theory at the Tachyonic Vacuum '' hep-th/0010240 } 
\lref\hkl{ Jeffrey A. Harvey, Per Kraus, Finn Larsen, 
  ``Exact non-commutative solitons,'' hep-th/0010060   } 
\lref\senuniq{A. Sen, `` Uniqueness of tachyonic solitons,'' hep-th/0009090 } 
\lref\blacad{ B. Blackadar, ``K-theory of operator algebras,''
 Springer-Verlag, 1986 } 
\lref\kk{ Hahimoto and Hashimoto } 
\lref\bp{ Bachas and Pioline } 
\lref\ter{ M. Hamanaka, S. Terashima, 
``On Exact Noncommutative BPS Solitons,'' hep-th/0010221. } 
\lref\hash{ K. Hashimoto, 
``Fluxons and Exact BPS Solitons in Non-Commutative Gauge Theory,''
hep-th/0010251. } 
\lref\furtop{ K. Furuuchi, 
``Topological charge of $U(1)$ instantons''hep-th/0010006  } 
\lref\furpeq{ K. Furuuchi,  
``Equivalence of Projections as Gauge Equivalence on Noncommutative
Space,'' hep-th/0005199 } 
\lref\sing{ ``Lectures on K-theory for physicists,'' I. Singer, MIT,
 FAll 2000 }  
\lref\grnk{ D. Gross and N. Nekrasov, ``Monopoles and strings in
 non-commutative gauge theory,'' hep-th/0005204 } 
\lref\grnki{ D. Gross and N. Nekrasov, ``Solitons in non-commutative 
 gauge theory,'' hep-th/0010090 } 
\lref\gms{ 
R.Gopakumar, S.Minwalla, and A.Strominger, ``Noncommutative solitons,'' {
  JHEP} {\bf 05} (2000) 020. } 
\lref\poly{ 
A.P. Polychronakos, ``Flux tube solutions in noncommutative gauge theories,''
 hep-th/0007043. }
\lref\bak{
D.Bak, ``Exact multi-vortex solutions in noncommutative Abelian-Higgs
  theory,''  hep-th/0008204.  } 
\lref\SW{N. Seiberg and E. Witten, ``String Theory and Noncommutative
Geometry'', JHEP {\bf 9909} (1999) 032.}
\lref\agms{ 
 M. Aganagic, R. Gopakumar, S. Minwalla, and A. Strominger,
 ``Unstable solitons   in noncommutative gauge theory,'' hep-th/0009142. } 
\lref\blp{ 
D.Bak, K.Lee, and J.-H. Park, ``Noncommutative vortex solitons,''
{ hep-th/0011099}. } 
\lref\muk{K. Dasgupta, G. Rajesh and S. Mukhi, 
``Noncommutative Tachyons,''  JHEP 0006:022,2000,   hep-th/0005006  } 
\lref\deal{S.P. de Alwis, ``Some issues in Non-commutative solitons,''
  hepth/0011223 } 
\lref\myers{ R.C. Myers, ``Dielectric Branes,'' hep-th/9910053, 
JHEP 9912 (1999) 022  }
\lref\wit{ E. Witten, ``D-branes and K theory'', JHEP {\bf 9812} (1998)
025.  } 
\lref\hamo{ J. Harvey and G. Moore, ``Non-commutative Tachyons 
 and K-theory,'' hep-th/0009030.  }
\lref\witov{E. Witten, ``Overview of K theory applied to strings'',
hep-th/0007175. }  
\lref\horava{ 
P. Horava, 
``Type IIA D-Branes, K-Theory, and Matrix Theory,'' hep-th/9812135 } 
\lref\minmo{ R. Minasian and G. Moore, 
``K Theory and  Ramond-Ramond charge,'' JHEP 9711:002,1997.  
  hep-th/9710230  } 
\lref\dgi{ N. Drukker, D. Gross and N. Itzhaki, ``Sphalerons, merons 
 and unstable branes in ADS,''  Phys.Rev.D62:086007,2000; 
hep-th/0004131 } 
\lref\hait{ A. Hashimoto and N. Itzhaki, 
``Traveling faster than  the speed of light in non-commutative geometry,'' 
hep-th/0012093. } 
\lref\nksh{  Nikita Nekrasov, Albert Schwarz, 
``Instantons on noncommutative $R^4$ and $(2,0)$ superconformal 
six-dimensional theory,'' 
Commun.Math.Phys.198:689-703,1998, 
hep-th/9802068  } 
\lref\ms{V. Mathai, I. Singer, 
``Twisted K Homology Theory, Twisted Ext - Theory, '' 
hep-th/0012046 }
\lref\ho{P. M.Ho, 
``Twisted Bundle on Noncommutative Space and $U(1)$ instanton,'' 
hep-th/0003012. }

\noblackbox

\def\Sd{ S^{\dagger } } 
\def\cM{ \cal{M} } 
\def\cA{ \cal{A} } 
\vskip 1cm

 \Title{ \vbox{\baselineskip12pt\hbox{  Brown Het-1250 }}}
 {\vbox{
\centerline{ Projector Equivalences in K theory and  }
\centerline{ Families of Non-commutative Solitons  }  }}

\centerline{$\quad$ {Steve Corley and Sanjaye Ramgoolam }}
\smallskip
\centerline{{\sl  }}
\centerline{{\sl Brown  University}}
\centerline{{\sl Providence, RI 02912 }}
\centerline{{\tt scorley, ramgosk@het.brown.edu}}
 \vskip .3in

 Projector equivalences used in the 
 definition of the K-theory of operator 
 algebras are shown to lead to generalizations 
 of the solution generating technique for  
 solitons in NC field theories, which has recently been 
 used in the construction of branes from other branes 
 in B-field backgrounds and in the construction of fluxon solutions
 of gauge theories.  The  generalizations involve 
 families of static solutions as well as solutions 
 which depend on euclidean time and interpolate between 
 different configurations. 
 We investigate the physics of these generalizations
 in the brane-construction as well as the fluxon context.
 These results can be interpreted in the light 
 of recent discussions on the topology of the configuration 
 space of string fields.  
 
%\draftmode 

\Date{12/2000 }

\newsec{ Introduction }
 
Solutions to non-commutative field theories 
on brane worldvolumes 
 have been of great interest recently
\refs{\nksh, \ho, \gms, \agms, \bak, \poly, \blp,
 \hait }. 
  Charges  of branes and the spacetime fields
 they couple to have  been shown to be closely 
 related to K-theory \refs{\wit\witov\minmo}.  
 In this paper we explore  applications 
  of some standard techniques in the K-theory 
 of operator algebras in the context of non-commutative 
 solitons. While the technical relations between the 
 two that we develop are  straightforward and of physical 
 interest, a bigger  picture  which would 
 explain these relations is not entirely clear.

A useful method for generating new solutions to noncommutative
field theories (NCFT) from old ones has recently been 
exploited in \hkl\ to construct lower dimensional brane solutions from known 
vacuum solutions in string theory.  The technique however is
quite general and has been applied in many other contexts as well,
see eg. \refs{\hkl, \hash, \ter, \deal}.
The idea is most easily formulated in the matrix or Hilbert
space representation of NCFT's discussed in \gms.  Conjugation of 
the fields in the
theory by some operator in the noncommutative Hilbert space
$U$ as in eg.
\eqn\phitran{\phi \rightarrow U \phi U^{\dagger}}
implies that the action $S$ transforms as
\eqn\actiontrans{S= \int {\rm Tr} {\cal L} \rightarrow
S_{U} = \int {\rm Tr} (U {\cal L} U^{\dagger})}
assuming that $U^{\dagger} U =1$.  The trace appearing in 
the action is over the noncommutative Hilbert space as well
as any gauge indices.  If $U$ can be cycled through the trace
then this transformation is a symmetry of the action.  However
this need not be so, in which case the transformation is not
a symmetry of the action but nevertheless remains a symmetry of
the equations of motion, i.e.,
\eqn\eomsym{{\delta S \over \delta \phi} = 0 \rightarrow
U {\delta S \over \delta \phi} U^{\dagger} = 0.}
Therefore from one solution follows another by conjugation.
Actually this holds whether $U$ generates a symmetry of the
action or just of the equations of motion.  In the former
case however $U$ would correspond to a (local or global)
gauge transformation so that the ``new'' solution would 
either be equivalent to the old one or simply related by
a global symmetry transformation.

When $U$ cannot be cycled through the trace it is clear that
the old and new solutions are not equivalent in general because
their energies differ.  This follows simply by noting that
the energy functional transforms in the same way as the
action functional above and therefore is not invariant.

An important point about the solution generating
technique is that it holds regardless of the value
of $U U^{\dagger}$, i.e., $U$ need not be unitary.
In fact all the solutions generated by this technique \refs{\hkl,
\hash, \ter} 
have used $U$'s which are not unitary but rather which satisfy
$U U^{\dagger} = P$ where $P$ is a projection operator. 
In the K-theory of operator algebras, projection operators
$P$ and $Q$ which are related by
\eqn\vnequiv{U^{\dagger} U = P, \, \, U U^{\dagger} = Q}
are known as von Neumann or algebraically equivalent.
It is a straightforward exercise, that we review in section
2, to show that these operators 
 become unitarily equivalent when
 embedded as projectors 
 in  $ {\cM}_2 ( \cA ) $, 
  the algebra of  $ 2 \times 2 $ matrices with entries in the algebra
 $\cA $.
  Further, embedding in $ {\cM}_{4} (A) $, we can 
  get a one-parameter family of unitary operators 
  $W(t)$ such that conjugating the embedded operator 
 $P$ leads to a family of projectors which interpolate
 from the embedded $P$ to the embedded $Q$. 
 We review these facts in the appendix and 
 comment on  their significance in the $K$ theory 
 of operator algebras. 

 In physical applications, the embedding into 
 ${\cM}_4( \cA)$ becomes relevant when we 
 go from a system with $U(1)$ gauge symmetry 
 to one with $U(4)$ gauge symmetry. This involves 
 generalizing a  system with a single brane to one 
 involving four branes. The family of interpolating projectors 
 can be used to construct a family of  static 
 solutions which interpolate between the trivial 
 solution and the new solution. By mapping the interpolating 
 parameter to the time variable in Euclidean space, we 
 can also construct time dependent solutions to the 
 Euclidean equations of motion. We consider two classes 
 of such solutions. The first interpolates the 
 solution based on $P$ to the  solution based on 
 $Q$. The second interpolates from the solution based
 on $P$ back to that based on $P$ after passing through the 
 solution $Q$. 

To get some insight into the physical significance of these 
 solution generating techniques we  
 consider two classes of examples. 
 The first involves solitons in unstable brane systems 
 which  are used to construct lower dimensional branes
 \refs{ \hklm, \muk, \hkl, \deal } 
  The second 
 involves fluxons, which are limits of non-commutative 
 monopole solutions \grnk.

\newsec{ Background  on non-commutative field theories. }

We begin by reviewing our conventions.
The noncommutative coordinates satisfy the commutator
conditions
\eqn\commcond{[\hat{x}^{i}, \hat{x}^j] = i \theta^{ij}.}
$\theta^{ij}$ can be block diagonalized into
$2 \times 2$ matrices.  Suppose this has been done and
consider the $i,j = 1,2$ block and define $\theta^{12} = \theta$.
Define the complex noncommutative coordinates 
\eqn\zzbar{ z:= {\hat{x}^{1} + i \hat{x}^{2} \over \sqrt{2}}, \, \,
\bar{z} := {\hat{x}^{1} - i \hat{x}^{2} \over \sqrt{2}}.}
These coordinates satisfy the commutation relation
\eqn\zcomm{[ z, \bar{z}] = \theta}
so that the annihilation and creation operators defined
as
\eqn\anncre{a := {z \over \sqrt{\theta}}, \, \,
a^{\dagger} := {\bar{z} \over \sqrt{\theta}}}
satisfy the usual commutation relation $[a, a^{\dagger}]=1$.
The algebra of annihilation and creation operators
can be realized on a Hilbert space in the
standard way, i.e., let
\eqn\Fock{{\cal H} = \bigoplus_{n=0}^{\infty} C |n \rangle}
where the action of $a$ and $a^{\dagger}$ on the $|n \rangle$
state is given by
\eqn\rep{a |n \rangle = \sqrt{n} |n-1 \rangle,
a^{\dagger} |n \rangle = \sqrt{n+1} |n+1 \rangle.}

Another representation of the algebra
is constructed in terms of $c$-number functions of the $x^{i}$'s by 
deforming the multiplication operation to the star product
$\star$ defined as
\eqn\definestar{f \star g (x) := \exp[{i \over 2} \theta^{ij} {\partial
\over \partial x^{i}_{1}} {\partial \over \partial x^{j}_{2}}]
f(x_{1}) g(x_{2}) |_{x_{1} = x_{2} = x}.}
One can map this representation into the above Fock space
representation by
the Weyl ordering map which takes a function of the
$c$-number coordinates $x^{i}$ into the associated
operator via 
\eqn\Weyl{\hat{f}(a,a^{\dagger}) = \int {d^{2}p \over 2 \pi}
e^{i(\bar{p} a + p a^{\dagger})} \tilde{f}(p)}
where
\eqn\Fourier{\tilde{f}(p) = \int {d^{2}z \over 2 \pi} 
e^{-i(\bar{p} z + p \bar{z})} f(z,\bar{z}).}
One property of this map is that
\eqn\inttrace{\Tr_{{\cal H}} \hat{f}(a, a^{\dagger}) =
{1 \over 2 \pi \theta} \int d^{2} z f(z,\bar{z}).}

To construct field theories over noncommutative spaces we need
derivative operators.  Derivatives generate infinitesimal
translations of the coordinates 
\eqn\trans{\delta_{\epsilon} x^{i} = \epsilon^{j} \partial_{j} x^{i} =
\epsilon^{i} .}
For noncommutative coordinates however this same action is generated
by commuting $\hat{x}^{i}$ with $-i \epsilon^{j} \theta_{jk} \hat{x}^{k}$.
Derivatives with respect to noncommutative coordinates can therefore
be written as
\eqn\deriv{\hat{\partial}_{i} f(\hat{x}) = [-i \theta_{ij} 
\hat{x}^{j}, f(\hat{x})].}
Derivatives with respect to $z$ and $\bar{z}$ are then transformed
into commutators with $a^{\dagger}$ and $a$ respectively via the
relations
\eqn\part{{\partial \over \partial z} = - {1 \over \sqrt{\theta}}
[a^{\dagger}, \cdot], \, \,
{\partial \over \partial \bar{z}} = {1 \over \sqrt{\theta}}
[a, \cdot].}

\newsec{ Solution generating operators, 
Unitary equivalence and homotopy } 

 Exact solutions are constructed by 
 conjugating a known solution by an operator $U$ satisfying
 \eqn\Ureln{U^{\dagger} U =1, \, \, U U^{\dagger} = P}
 where $P$ is a projection operator.  A particularly simple
 example is given by the shift operator 
 $S = \sum_{i=0}^{\infty}  |i+1 \rangle \langle i  | $ which satisfies 
 the condition 
\eqn\conds{\eqalign{&  S^{\dagger }S = 1 \cr 
                    & S S^{\dagger}  = 1 - P_0 }}
where $P_{0} = |0 \rangle \langle 0|$.
Acting on  operators as $ {\cal O}  \rightarrow  S { \cal O } \Sd $
is a symmetry of the equations of motion 
 and allows us to generate new solutions 
 from old ones. It is however not a symmetry 
 of the action because $S$ cannot be cycled through the trace as
 discussed in the introduction.
 In general the condition for being able to cycle operators
 is  ${\rm Tr} (AB ) = {\rm Tr} ( BA ) $ if $ A$ is trace-class and 
 $B$ is bounded. For the shift operator these conditions are 
 not satisfied and we get a different action, and futhermore a different
 energy.

%The index of the operator is an integer which measures 
% the difference in rank of the above two things. ( REF for discussions
% of this ) 
 
The equations \conds\ can be expressed as establishing that 
 the identity operator and $ 1- P_0$ are algebraically
  or Von-Neumann equivalent as projectors ( see appendix for more 
 on this ).  
Another notion of equivalence of projectors is given by unitary equivalence.
In general Von Neumann equivalent projectors will not be unitarily
equivalent.  If however von Neumann equivalent
projectors are embedded in higher dimensional
matrices they can be made unitarily equivalent.  In the $2 \times 2$
case for the shift operator example given above we embed the 
projectors $I$ and $I-P_{0}$ as
\eqn\embed{\eqalign{  I \rightarrow \pmatrix { &  I & 0 \cr 
                        &  0 & 0 }, \, \, 
 I-P_{0} \rightarrow \pmatrix { &  I-P_{0} & 0 \cr 
                        &  0 & 0 }. }}
It is then straightforward to check that these matrices are
related by the unitary transformation  
\eqn\relun{\eqalign{  Z = \pmatrix { &  S & P_0 \cr 
                        &  0 & S^{\dagger} } } }
with inverse
\eqn\inv{\eqalign{  Z^{-1} = \pmatrix { &  S^{\dagger} & 0  \cr 
                        &  P_0  & S }. }}
Note that $S$ and therefore $Z$ are not trace class as follows
simply from the fact that  
the sum of eigenvalues of $ \sqrt { S^{\dagger }S  } $ is infinite. 
 
 A third notion of projector equivalence, and the one that
 we shall primarily exploit in this paper, is homotopy equivalence.
 For our purposes homotopy equivalence means that there exists
 a one parameter family of projection operators which interpolate
 between the two given projectors.  Homotopy equivalence of
 projectors implies unitary equivalence, but not vice versa.
 Unitary equivalence in an algebra 
 ${\cal A}$ however implies homotopy equivalence when embedded
 in $ {\cal M}_2({\cal A}) $. 
 Suppose $E$ and $F$ are unitary equivalent projectors
 satisfying $ E = Z F Z^{-1} $, then there is a family 
 of operators  interpolating between 
 $$ \hat E =  \pmatrix{ & E & 0 \cr 
              & 0 & 0}  $$ 
 and $$ \hat F  = \pmatrix{ & F & 0 \cr 
              & 0 & 0}. $$ 
 The interpolating matrix is 
 \eqn\homotopy{\eqalign{
W_t =  \pmatrix{ &  ( c^2 + Z s^2)  & cs ( 1 - Z ) \cr 
                   & cs ( Z^{-1} - 1 ) & c^2 + Z^{-1} s^2 \cr } }} 
 where $ c = {\rm cos} ( {\pi t \over 2} ) $ and $s = 
{\rm sin} ( {\pi t \over 2})$.  In particular $W_{t}$ at $t=0$
is the identity and at $t=1$ is  $ W = \pmatrix{ & Z & 0 \cr 
                            & 0 & Z^{-1}} $.
Since $Z$ is unitary it follows that $W_t$ is also unitary and its inverse
is given by  
\eqn\homtinv{\eqalign{ 
W_t^{-1}  =  \pmatrix{ &  ( c^2 + Z^{-1}  s^2)  & cs (  Z - 1   ) \cr 
                  & cs ( 1- Z^{-1}  ) & c^2 + Z s^2 \cr }. }}  

 Note in our application $E$,  $F$, $Z$  are themselves 
 $2 \times 2$ matrices with entries which are operators 
 in the Hilbert space of a free oscillator. The expression for
 $Z$ is given above in \relun. 
 Therefore,  $ \hat E $ and $\hat F$  are $ 4 \times 4$ matrices 
 with entries which are operators in the Hilbert space. 
 More explicitly the matrix is 
\eqn\hmtpy{\eqalign{  
W_t =  \pmatrix{ &   c^2 + S  s^2 & P_0 s^2 &    cs ( 1 - S )  & - cs P_0 \cr
                &   0 & c^2 + s^2 S^{\dagger}  &    0&
   cs ( 1-S^{\dagger})  \cr  
                   & cs (S^{\dagger} - 1 )& 0 & c^2 + s^2 S^{\dagger}
 & 0 \cr 
                   & cs ( P_0 ) & cs (S -1 ) & s^2 P_0 & c^2 + s^2 P_0
 \cr  }.
}}

% The upshot is that given the fact that $S$ 
% generates a solution of the $U(1)$ theory, 
% and choosing  a trivial embedding of the 
% solution as a solution of the $U(4) $ theory, 
% we can find a family of solutions  which  are 
% continuously interpolate between the trivial solution and the 
% embedded fluxon. 
% When we embed in $U(2)$ theory we have unitary equivalence, 
% but not homotopy equivalence. The unitary equivalence 
% means that the vacuum and  the solution are related by an operatr 
% of zero index. 

%EXERC: Describe the configurations that interpolate between 
% vacuum and $U(4)$ embedded fluxon in terms of brane configurations. 

\newsec{ Non-commutative tachyons } 

We now use the formalism discussed in the previous section to
construct new solutions to the equations of motion from known
ones in various settings.  Of particular interest will be
the homotopy equivalence of
projection operators which leads to new  solutions which interpolate 
between simple embeddings of recently constructed 
solutions into higher rank gauge groups. 
In this section we will  consider solutions 
of unstable brane worldvolumes in the presence of 
B-fields,  which represent lower branes.  
In the next section 
we will consider solutions of three-brane worldvolume theory
 which represent geometrical worldvolume 
deformations  associated with D-string charge ( fluxons ). 
We will find that some of the results in this section 
 shed light on the topology of string configuration space.

\subsec{ $D25$ in Bosonic string and $D9$ in Type IIA  }

 The discussion in these two cases is almost identical. 
 We will for simplicity discuss the construction 
 of $D23$ in $D25$. Substituting unstable $D7$
 inside unstable $D9$ of Type IIA is a trivial generalization. 
 
 Following the notation of \hkl\ the bosonic string action 
for the tachyon and gauge
field with vanishing $B$ field is given by  
 \eqn\zda{S = {c \over g_s} \int\! d^{26}x \sqrt{g}\left\{
-{1 \over 4}h(\phi-1)F^{\mu\nu}F_{\mu\nu}   + \cdots +     
{1 \over 2}f(\phi-1)\partial^\mu \phi \partial_\mu \phi + \cdots - V(\phi-1)\right\}}
where
\eqn\zfe{F_{\mu\nu} = \partial_\mu A_\nu - \partial_\nu A_\mu 
+i[A_\mu,A_\nu]~.}
In this notation the closed string vacuum corresponds to $\phi = 1$
at which point the potential has a local minimum and
vanishes along with $h$, i.e., $V(0) = 0 = h(0)$.  The open 
string vacuum corresponds to $\phi = 0$ where the potential
has a local maximum and has been normalized as $V(-1) = 1$.

The prescription for generalizing the action \zda\ to a 
constant $B$ field background has been given in \SW.
Specifically
all products in the action are deformed
to star products (which then become
matrix products in the Hilbert space representation discussed in
section 1) with noncommutativity
parameter given in terms of $B$ and the closed
string coupling $g_s$ and closed string metric $g_{\mu \nu}$
are replaced by the corresponding open string quantities $G_s$ and 
$G_{\mu \nu}$\foot{The relations between these parameters
can be found in eg. \hkl. }
% We shall not need them below so
%do not bother to write them out
The end result is the action
\eqn\zff{\eqalign{S = {2 \pi \theta c \over  G_s} \int\! d^{24}x 
\sqrt{G}~{\rm Tr}
&
\left\{ 
-{1 \over 4}h(\phi-1)(F^{\mu\nu}+\Phi^{\mu\nu})(F_{\mu\nu}+\Phi_{\mu\nu})
   + \cdots 
\right. 
\cr
 &~~~+ 
\left.    
{1 \over 2}f(\phi-1)D^\mu \phi D_\mu \phi + \cdots - V(\phi-1) 
\right\}
~}}
with
\eqn\zfg{F_{24,25}+\Phi_{24,25} = -iF_{z\overline{z}} + {1 \over
\theta }
= -{1 \over \theta }[C,\bar C]~.}
There is some freedom in writing this action given by the
choice of $\Phi_{\mu \nu}$.  Following \hkl\ we have made the
choice given in \zfg.

In the U(N) case, $C$ becomes an $ N \times N $ matrix
whose entries are elements in the Hilbert space  
of a free oscillator. In this case the 
 leading terms in the action 
 can be rewritten 
 as 
 \eqn\zffi{\eqalign{S = {2 \pi \theta c \over  G_s} \int\! d^{24}x 
\sqrt{G}~{\rm Tr}
&
\left\{ 
- {1 \over 4} h(\phi - 1) F^{mn} F_{mn } - {1 \over 4 \theta^2 }
 h(\phi - 1) ( [ C, \bar C ] )^2 
    + \cdots 
\right. 
\cr
 &~~~+ 
\left.    
{1 \over 2}f(\phi-1)D^\mu \phi D_\mu \phi + \cdots - V(\phi-1) 
\right\}
~,}}
where we have used
\eqn\zfg{F_{24,25}+\Phi_{24,25} = -iF_{z\overline{z}} + {1 \over \theta}
= -{1 \over \theta}[C,\bar C]~.}
The indices $m,n$ extend from $ 1 $ to $24 $ and
the trace now includes summing over the 
Hilbert space indices as well as the 
colour indices.

The solution generating transformation 
 acts in the $U(1)$  case as : 
\eqn\ze{\eqalign{\phi \rightarrow & ~U \phi \bar U~, \cr
C \rightarrow & ~U C \bar U~,\cr
A_\mu \rightarrow & ~U A_\mu \bar U~, \quad \mu =0 \ldots 23~. }}
Acting on the closed string vacuum 
$$ \phi = 1, C = \bar a, A_{\mu } = 0 $$ 
we obtain the solution
\eqn\oned{\eqalign{& \phi = S  \Sd = (1 - P_0) \cr 
                   &  C = S {\bar a}   \Sd \cr 
                   &  A_{m}= 0. \cr }}
This solution was identified in \hkl\ as the D23 brane
as its tension can be shown to be that of the D23
brane.

\subsec{ Interpolating between static solutions : General remarks }

 In the following we will apply the formalism developed in 
  section 3 to interpolate between 
 simple embeddings of $U(1)$ solutions into $U(4)$.
 Specifically we embed $U(1)$ solutions such as the
 closed and open string vacua into $U(4)$ in a trivial
 manner so that the $U(4)$ field equations are satisfied.
 We then conjugate these solutions by $W(t)$ constructed
 in section 2, thereby generating new $t$-dependent
 solutions.  A natural question to consider is the
 interpretation of the parameter $t$. 
 The most straightforward 
 interpretation is that it simply parametrizes
 a family of static classical solutions.
 In this sense the solutions are like a family 
 of sphalerons parametrized by $t$  ( for recent
  discussions of these in string theory see \refs{\hhk, \dgi  }). 
 They are not exactly sphalerons because 
 they do not have finite energy. 
 The infinities in the energy are however 
 easy to understand and can be regulated 
 by working on a non-commutative $T^2$ rather 
 than a non-commutative $R^2$. 
 The parameter $t$ just labels 
 a static solution which lies at an intermediate  
 point between the two solutions we are interpolating between. 

 Another possible interpretation of $t$ is as Euclidean time.
 Specifically the solution generating transformation
 also allows us to construct solutions 
 to the Euclidean space-time equations 
 of motion by mapping the interpolating parameter 
 $t$ to the Euclidean time $x_0$. There is a moduli 
 space corresponding to choices 
 of the function $x_0(t) $ subject to some 
 boundary conditions. In this case we also conjugate 
 the time-covariant derivative, 
  \eqn\trnsdt{ D_0 \rightarrow W(x_0) \partial_t W(x_0)^{-1}. } 
 The  first description above is just a special 
 case of this moduli space where $ x_0$ is chosen to be independent 
 of $t$.  
We will see that  the action of these solutions    
 is infinite. In some case this infinity 
 can be expected on physical grounds. 
 In other cases, it is conceivable 
 that some deformation of these solutions 
 can give finite action ( although we do not 
 have any concrete directions for the right deformations 
 at this point).

\subsec{ Case I } 
Consider the following solution 
to the equations of motion of the bosonic string : 
\eqn\stconfii{\eqalign{ 
& \Phi^{(0)}  = \pmatrix{ & 1 & 0 & 0 & 0 \cr 
                         & 0 & 0 & 0 & 0 \cr 
                         & 0 & 0 & 0 & 0 \cr 
                         & 0 & 0 & 0 & 0 \cr } \cr 
& C = \bar a \pmatrix{ &  1 & 0 & 0 & 0 \cr 
                       & 0 & 0 & 0 & 0 \cr 
                       & 0 & 0 & 0   & 0 \cr 
                       & 0 & 0 & 0  &   0\cr } \cr }}

After applying the conjugating transformation 
\eqn\cts{ U = 
 \pmatrix{ & S & 0 & 0 & 0 \cr 
                         & 0 & 0 & 0 & 0 \cr 
                         & 0 & 0 & 0 & 0 \cr 
                         & 0 & 0 & 0 & 0 \cr } }
the configurations we get are: 
\eqn\ndconfii{ \eqalign{ 
&  \Phi  = \pmatrix{         & 1 - P_0 & 0 & 0 & 0 \cr 
                                   & 0       & 0    & 0 & 0 \cr 
                                   & 0       & 0    & 0 & 0 \cr 
                                   & 0       & 0    & 0 & 0 \cr } \cr 
&   C =  \pmatrix{ &  S \bar a \Sd & 0 & 0 & 0 \cr 
                       & 0 & 0 & 0 & 0 \cr 
                       & 0 & 0 & 0 & 0 \cr 
                       & 0 & 0 & 0  &  0  \cr } \cr }}
The interpolating configurations are 
\eqn\icfiii{\eqalign{
 & \Phi ( t ) = W_t \Phi^{(0)} W_t^{-1}  \cr  
           & = \pmatrix{ c^4 + c^2s^2( S + \Sd ) + s^4 ( 1 - P_0 ) & 0
& -c^3s ( 1 - S ) - cs^3 ( 1 - S ) S & c^3s P_0 + cs^3 S P_0 \cr 
 0 & 0 & 0 & 0 \cr 
 c^3s ( \Sd - 1 ) + cs^3 \Sd ( \Sd - 1 ) & 0 & -c^2s^2( S + \Sd -2 )
& -c^2s^2 P_0 \cr 
 c^3s P_0 + cs^3 P_0 \Sd & 0 & -c^2s^2P_0 & c^2s^2 P_0  \cr } }}
and 
\eqn\icfiiig{\eqalign{ 
&C   = W_t C^{(0)} W_t^{-1}  \cr 
   & = \pmatrix{ c^4 \bar a  + c^2s^2( S \bar a + \bar a \Sd ) +
 s^4 S \bar a \Sd  & 0
& -c^3s \bar a ( 1 - S ) - cs^3 ( S \bar a  -   S \bar a   S ) & c^3s
\bar a  P_0 + cs^3 S \bar a  P_0 \cr 
 0 & 0 & 0 & 0 \cr 
 c^3s ( \Sd  - 1 ) \bar a  + cs^3 ( \Sd  \bar a  \Sd - \bar a  )
 & 0 & c^2s^2 ( \Sd - 1  ) \bar a ( S -1 ) 
& c^2s^2  ( \Sd - 1 ) \bar a  P_0 \cr 
 c^3s P_0 \bar a  + cs^3 P_0 \bar a \Sd 
 & 0 & c^2s^2 P_0 \bar a ( S - 1) & c^2 s^2 P_0 \bar a
P_0  \cr } }}
where \ndconfii\ is reproduced at $t=1$.

 For the moment we interpret $t$ as the parameter in a one
 parameter family of static solutions and now compute the
 energy of the interpolating configurations.
 The contribution to the energy from the gauge field 
 is proportional to the quantity 
\eqn\cont{\eqalign{ 
  [ C, \bar C ]^2(t) &= W(t) [C, \bar C ]^2 (0) W(t)^{-1} \cr 
                     &= W(t) Diag ( 1, 0, 0, 0 ) W(t)^{-1} \cr 
                     & = \Phi ( t) \cr }}
The coefficient of proportionality is  
 $ h( \Phi(t) -1 ) $, multiplying the two yields
\eqn\cont{ h ( \Phi(t)  - 1 ) [ C, \bar C ]^2(t) 
           = h (-1)  ( -\Phi(t)  + 1 )  \Phi ( t) =0.} 
In the last step we have used the fact that 
 $ \Phi(t) $ is a projector, and $ h(0) = 0 $.
 Alternatively we can arrive at the same 
 result by noting that $h( \Phi (0)  - 1 ) [ C, \bar C ]^2(0)$ 
 is zero since $h$ is zero in the first block and 
 the field strength is zero in the other blocks. 
 This quantity at time $t$ is obtained by conjugating
 with $W(t)$, so this remains zero.

%The time derivatives of $\Phi $ also give zero
%since $ D_0 \Phi D_0 \Phi $ is obtained by conjugating 
%the covariant derivatives at $ t=0 $ which are zero. 

Since the solution is static there is furthermore no contribution
to the energy from time derivatives.  
The only non-zero contribution to the energy therefore comes
from the potential and is 
\eqn\nz{\eqalign{ 
 \int d^{23 } x  {\rm Tr} V ( \Phi -1 ) 
&= \int d^{23} x \int V (-1) {\rm Tr}  (1- \Phi )   \cr  
                       & = \int d^{23} x (3 {\rm Tr}(1)
 - {1 \over 2} sin^2 ( { \pi t \over 2 }) cos ( \pi t )) \cr }} 
The first term can be interpreted as the energy of 
 the three $D25 $ branes. The second term vanishes
 at $t=0$ corresponding to the closed string vacuum
 and at $t=1$ gives the energy of the $D23$ brane.
 At intermediate values of $t$ the extra contribution
 to the energy can be either positive or negative.

 The parameter $t$ can also be interpreted as Euclidean
 time.  In this case while the 
 fields $\Phi, C $ indeed take the values 
 at  $ t=0 $ given in equation 
 \stconfii\ and end at the configuration 
 \ndconfii\ at $t=1$ the time derivative
 of $ \Phi $ is non-zero, and in particular non-zero at
 $t=0$ and $t=1$.  Specifically
\eqn\tmder{
 \partial_t \Phi = [ ( \partial_t W) W^{-1} , \Phi ] } 
which gives 
\eqn\tmderi{\eqalign{ 
 \partial_t \Phi  (0) &= \pmatrix{&  0 &  0 &  { 1\over 2 }( S-1) & {1
\over 2 } P_0 \cr 
                                 & 0 & 0 & 0 & 0 \cr 
                                 & { 1 \over 2 } ( S^{\dagger} - 1 ) &
                                 0 & 0 & 0 \cr 
                                 & { P_0\over 2 } & 0 & 0 & 0 \cr }
                                 \cr 
\partial_t \Phi  (1) &= \pmatrix{&  0 &  0 &    
                            {1\over 2 }S ( S-1) &  ( 1 - S )
                                 { P_0\over 2 }  \cr 
                                 & 0 & 0 & 0 & 0 \cr 
                                 & { 1 \over 2 } \Sd ( S^{\dagger} - 1 ) &
                                 0 & 0 & 0 \cr 
                                 & { P_0\over 2 }( 1 - \Sd )  & 0 & 0
                                 & 0
 \cr } \cr }}
The covariant derivative is of course zero. 

It is possible to construct solutions 
where the time derivatives at the beginning and end points 
 are zero. This is achieved by essentially 
 performing a redefinition of the time variable 
 such that the interval $[0,1 ]$ is streched to 
 $ [ - \infty , \infty ]$.  For example we can let Euclidean 
 time $x_0$ be a function of the interpolating 
 parameter $t$ as $x_0 = \tan \pi( t - {1\over 2 } )  $
 The appropriate solution is 
 obtained by conjugating with a matrix 
 $ U ( x_0 ) = W( t (x_0) )  $. 
  
A natural quantity to consider for Euclidean solutions
is the value of the action.  For the solution constructed
above this is easy to compute using our previous computations
for the energy of the one parameter family of static solutions.
In particular the only non-zero contribution again comes from
the potential $V$ giving us
\eqn\instact{ S = {2 \pi \theta c \over G_{s}} \int d^{23} x dx_{0}
{\rm Tr} V(\Phi(t(x_{0})) - 1)}
where $\int d^{23} x V(\Phi(t)-1)$ is given in \nz.  
It follows that the action in this case will contain
the usual divergence from the $d^{23}x$ integral
as well as the ${\rm Tr} ~ (1)$ divergence.  Both
could be regulated by working on compact commutative
and noncommutative spaces respectively.  The action
however contains another
divergent contribution coming from the $x_{0}$ integral.
For the ${\rm Tr} ~ (1)$ term this follows because it is
$x_{0}$ independent while the integral of the remaining $x_{0}$ dependent
term diverges at the $x_{0} \rightarrow \infty$ limit.
%This 
%is an infinity which goes beyond what may be expected 
%to be regulated by working on finite volumes making an
%instanton interpretation of this Euclidean solution 
%difficult. 
 
%The infinite contribution proportional to the 
%$D23$ brane action can be made finite if we interpolate 
% from the configuration 
%at $ t=0 $ back to itself by letting 
%$ t $ run from $ 0$ to $2$ and mapping that
% to infinity. 

The divergent $x_{0}$ integral can be partially cured
by instead letting $t$ run from 0 to 2 and stretching this
interval to infinity by eg. the map
$ x_0 = tan {\pi\over 2 }( t - 1 )  $.  
The Euclidean time integral of the  $x_{0}$ dependent term in the 
potential, the second term in \nz,  now becomes finite.
 There is still an infinite $ \int dx_0 $ multiplying $ { \rm Tr} ~ (1) $. 
This solution corresponds to `evolving' in $x_{0}$
from the vacuum configuration \stconfii through the
solution \ndconfii and then back to \stconfii.
 It can further be modified to ensure 
 that the time derivative $ \partial_{0}$ 
  is zero when the configuration passes through 
 \ndconfii\ at $t=1$. 
There is a large class of such solutions 
interpolating from one vacuum to the same
parametrized by a choice of finite function 
living on the interval $ [0,2]$.
These solutions are in some sense like an instanton-anti-instanton
pair.  The endpoints of the solution are the mixed closed
string vacuum, open string vacuum configurations given in
\stconfii.  At some finite $x_{0}$ the configuration consists
of an unstable $D23$ brane and the open string vacuum as in
\ndconfii.  Therefore at least for these configurations the
open string vacua appearing in the solution are basically 
just spectators with the evolution all occuring in one
particular $U(1)$ sector.  This action also splits 
into a sum of terms associated with each block.
  The ${\rm Tr}~ (1)$ part corresponds to the
three $D25$ branes in their vacuum states while the remaining
finite contribution corresponds to the closed string vacuum -
D23 brane - closed string vacuum sector.  Of course at other values
of $x_{0}$ the solution $\Phi(t(x_{0}))$ has off-diagonal
elements so it is difficult to make this interpretation
rigorous.   
%These solutions can be interpreted as an ``instanton-anti-instanton''
%pair. It starts off at $t=0$ ( $ x_0 = - \infty$) as  a 
%configuration without 
%lower brane charge, passes through the  configuration 
% with lower brane \ndconfii\  at $t=1$ ( $ x_0 $ finite  )  
% and goes back to the original configuration at 
%$ t = 2 $ ( $x_0 = \infty $ ). 
Nevertheless it is rather surprising that 
such ``instanton-anti-instanton'' like 
 configurations exist as exact solutions.

\subsec{ Some other interpolating solutions } 
 
 If we start with a general diagonal configuration 
 of fields which can be interpreted simply, 
 we are not guaranteed to end with a diagonal 
 configuration at $t=1$ after conjugation by 
 $W$. Two more examples where this is possible 
 are given as follows. 
 If we start with $ \Phi = Diag ( 0 ,  1 -P_0 ,  1,  1  ) $,  
 we also end up with something diagonal
  $ \Phi = Diag ( 0, 1, 1, 1 ) $.   
  Another pair of diagonal solutions 
 we can interpolate between is 
$ \Phi = Diag ( 1,1,1, 1 ), C = \bar a Diag ( 1,0,0,0 ) $ 
and $  \Phi = Diag ( 1,1,1, 1 ), C = S \bar a \Sd  Diag ( 1,0,0,0 ) $

\subsec{ $D-\bar D$ system, Elementary strings and a puzzle } 

For the superstring, unlike the bosonic string, the D-branes
carry conserved charges from the Ramond-Ramond sector.  Furthermore,
unlike with the bosonic string, single D-brane configurations are
stable.  However if a D-brane and it's oppositely charge version,
the ${\bar D}$-brane, are brought together, they can annihilate.
This instability manifests itself from the worldsheet point of view
through a complex tachyon field which is charged under the 
$U(1) \times U(1)$ gauge symmetry of the D($\bar{\rm D}$)-brane
respectively, i.e., the tachyon transforms in the bifundamental
of the $U(1) \times U(1)$ gauge symmetry.  In the noncommutative
case then it follows that $\phi$ transforms as
\eqn\tactran{\eqalign{& \phi \rightarrow U \phi {\bar V} \cr
& \bar{\phi} \rightarrow V \bar{\phi} \bar{U}.}}
under a gauge
transformation, and more generally under the solution generating
transformation.  Similarly the gauge fields in the noncommutative
directions transform as
\eqn\gaugetrans{\eqalign{& C^{+} \rightarrow U C^{+} U^{\dagger} \cr
& C^{-} \rightarrow V C^{-} V^{\dagger}.}}

This leads to some rather unexpected phenomena when we
embed in $U(4)$ as before.  In particular it appears that
we can construct an interpolating solution between any two
$D7 - {\bar D7}$ configurations.  To see this take the
following field configuration for the tachyon and gauge
fields in the noncommutative directions 
\eqn\mixedvac{\eqalign{& \Phi = {\rm diag}(1,0,0,0) \cr
& C^{+} = {\rm diag} (a^{\dagger},0,0,0) \cr
& C^{-} = {\rm diag} (a^{\dagger},0,0,0).}}
Now conjugate this `ground' state by
\eqn\massiveconj{U = (W_{t})^{n}, \,\, V=(W_{t})^{m}.}
The time dependent tachyon becomes
$\phi(t) = (W_{t})^{n} \phi (W^{\dagger}_{t})^{m}$, and similarly
for the gauge fields.  Evaluating at $t=1$ we find that
\eqn\arbit{\eqalign{& \phi(1) = {\rm diag}(S^{n} (S^{\dagger})^{m},0,0,0) \cr
& C^{+}(1) = {\rm diag}(S^{n} a^{\dagger} (S^{\dagger})^{n},0,0,0) \cr
& C^{-}(1) = {\rm diag}(S^{m} a^{\dagger} (S^{\dagger})^{m},0,0,0).}} 
The (1,1) component of these fields is exactly the $n-D7, m-{\bar D7}$
configuration identified by \hkl. Furthermore it is easy to check that
by applying $U= W^{\dagger}_{t}$ and $V=1$ or $U=1$ and $V=W^{\dagger}_{t}$ 
one can remove
the number of $D7$ and ${\bar D7}$ branes respectively.
This is not what we would have
naively expected to be the case.

%\eqn\sol{\eqalign{ &  \Phi = \pmatrix{ & 1 & 0 & 0 & 0 \cr 
%                           & 0 & 0 & 0 & 0 \cr 
%                           & 0 & 0 & 0 & 0  \cr 
%                           & 0 & 0 & 0 & 0 \cr }   \cr 
%                    & C  = \pmatrix{ &  \bar a  & 0 & 0 & 0 \cr 
%                                     & 0 & 0 & 0 & 0 \cr 
%                                     & 0 & 0 & 0 & 0 \cr 
%                                     & 0 & 0 & 0 & 0 \cr }
%}}

 The same constructions can be used to 
 construct closed string solitons in 
 non-commutative gauge theory.  
 Some relevant papers are 
 \hklm\senel. 
  Here the solutions involve 
  $ \Phi = 1 $ and $ E = P_0 $. 
  Consider the $U(4) $ configuration 
   $ \Phi =Diag  ( 1,1, 1, 1 ), E = (P_0, 1 , 0,0)   $ 
  The tachyon is sitting at the closed string 
  minimum here in all four blocks, but 
  the electric flux describes an elementary string 
  soliton in the first  block. The second block 
 may be interpreted as describing infinitely 
 many closed strings.  
 Conjugating by $ Diag (Z,Z^{-1})  $ we get 
  $ \Phi =Diag  ( 1,1, 1, 1 ), E = (0,1, 0,0)   $, i.e 
 we have got rid of the closed string in the first block. 
 By embedding further in $ U(7) $
 we can interpolate to a configuration 
 where the elemenatry string has been replaced 
 by an unstable brane of codimension $2$. 
Starting with 
$ \Phi =Diag  ( 1; 1, 1, 1 ; 0,0,0 ), E = (P_0; 1, 0,0 ; 0,0,0)   $
we can interpolate to
$ \Phi =Diag  ( 1; 1, 1, 1 ; 0,0,0 ), E = (0; 1, 0,0 ; 0,0,0)   $
 using matrices which act non-trivially 
 on the first and the second set of three 
 entries. 
Then using matrices acting on the first entry 
 and the second set of three entries
 we can get to 
 $ \Phi =Diag  ( 1- P_0 ; 1, 1, 1 ; 0,0,0 ), E = (P_0; 1, 0,0 ; 0,0,0)
 $. 
In this way we have interpolated, in the first block, 
 between an elementary string 
 soliton and an unstable brane by passing through the vacuum.

That we could interpolate between 
 vacuum configurations 
 and these configurations which 
 contain charged objects like a BPS D-brane 
 or a closed elementary string seems somewhat surprising. 
 These interpolating solutions are not finite 
 action instanton solutions so there is 
 no immediate contradiction involving 
 transitions from a direct sum of open and 
 closed string vacua  to a state 
 containing a charged brane, or transitions 
 from the vacuum to charged brane and back to the vacuum. 
 However even the special case of $ \partial_t x_0 = 0 $, 
 where we do not view these as solutions to Euclidean 
 equations of motion but rather as families of 
 on-shell configurations,  
 is intriguing. It shows that there is a family 
 of static solutions which connect the vacuum 
 to the brane configuration. 
 In the case where the brane being created
 is a D23-brane or an unstable 
 D7-brane as in the earlier subsection 4.2, 
 the existence of such configurations 
 is expected 
 from the discussion of the topology 
 of string configuration space where unstable branes 
 are interpreted as sphalerons \hhk. The fact that the family 
 of configurations is on-shell is not predicted 
 by that discussion, but its existence is not surprising. 
 However the interpolation in the case of charged 
 branes implies that the conservation of RR charge 
 is not related to the topology of string configuration 
 space but to some more subtle topology whose relation 
 to the topology of string configuration space 
 remains to be clarified. The existence of these 
 interpolations can again to some extent
 be anticipated by the considerations of \senuniq. 
 There it was shown, by CFT arguments,  that while magnetic flux on 
 $T^2$  labels different 
 sectors in the space of fields for Yang-Mills 
 on the torus, configurations with different flux can 
 be connected in  string configuration space. 
 In that case one needed the stringy description 
 to see the interpolation but here the fields of NC Yang-Mills
 suffice. 
 
 Another possible solution of the puzzle which we consider
 less likely, but are unable to dismiss completely, 
  is  that the naive interpretation of the 
 block diagonal configuration, which leads to the 
 conclusion that the charge of 
 the block diagonal configuration is just the 
 sum of charges associated with each block, 
 is missing some subtlety of the full non-abelian 
 brane world-volume action, and actually has zero  
 brane charge. One might be tempted to think 
 this is the case because the solutions can be  generated
 in the $U(4)$  case by conjugating the vacuum 
 with an operator of index $0$. Connections 
 between the index of the conjugating operator 
 and charge have been discussed in \grnk\furtop\hamo\   for
 example but it is  hard to see, in terms of brane actions 
 how the direct sum configuration could fail to have 
 the direct sum of charges. 

 Since we have seen that 
 this solution generating technique 
 has implications for the string 
 configuration space and leads to an interesting puzzle, 
 it is instructive to study its implications
 in a context where there are no tachyons and 
 the system is as simple is as possible. 
 In the following we look at the fluxons
 and show that sensible  interpolations 
 between $U(4)$ configurations which have a clear meaning 
 in terms of $D3$ and $D1$ branes is possible. 

\newsec{ Fluxons and Noncommutative gauge theory}

We review here a class of solutions of
${\cal N} = 4$ supersymmetric noncommutative $U(1)$
gauge theory discussed in  \grnk\grnki\hash\ter. 
The action  (with $x^{1}$ and $x^{2}$ the noncommutative
directions) is given by
\eqn\action{S = {2 \pi \theta \over g^2} \int dt dx^{3} \Tr [- {1 \over 4}
F_{\mu \nu} F^{\mu \nu} + {1 \over 2} D_{\mu} \Phi_{a} D^{\mu} 
\Phi_{a} - {1 \over 4} [\Phi_{a}, \Phi_{b}]^{2}] + {\rm fermions}} 
where the covariant derivative is given by
\eqn\covder{D_{\mu} = \partial_{\mu} - i A_{\mu}}
and the curvature tensor by
\eqn\curv{F_{\mu \nu} = i([D_{\mu},D_{\nu}] - i \theta_{\mu \nu}).}
The equations of motion following from this action are
\eqn\eom{\eqalign{[D^{\nu}, [D_{\nu}, D_{\mu}]] &  = [\Phi_{a},
[D_{\mu}, \Phi_{a}]] \cr
[D^{\mu}, [D_{\mu}, \Phi_{a}]] & = [\Phi_{b}, [\Phi_{b}, \Phi_{a}]].}}

Now consider static solutions with no electric potential
$A_{0} = 0$ and only one nontrivial scalar field $\Phi$.
Defining the magnetic field as
\eqn\mag{B_{i} := {1 \over 2} \epsilon_{ijk} F_{jk}}
one can show that the first order BPS equations
\eqn\BPS{B_{i} + [D_{i}, \Phi] = 0}
are consistent with the equations of motion.
Working in the gauge $A_{3} = 0$ the BPS equations reduce to
\eqn\reduced{\eqalign{{\partial \Phi \over \partial x^{3}} 
& = {1 \over \theta} ([C,\bar{C}] + 1) \cr
{\partial C \over \partial x^{3}} & = - [C, \Phi] \cr
{\partial \bar{C} \over \partial x^{3}} & = [\bar{C}, \Phi]}}
where we have defined $C$ and $\bar{C}$ as
\eqn\defineC{C := (a^{\dagger} + i \sqrt{\theta} A), \, \,
\bar{C} = (a - i \sqrt{\theta} \bar{A})}
with $A$ and $\bar{A}$ given in terms of $A_{1}$ and $A_{2}$ as
\eqn\defineA{A := {A_{1} - i A_{2} \over \sqrt{2}}, \, \,
\bar{A} := {A_{1} + i A_{2} \over \sqrt{2}}.}
$C$ and $\bar{C}$ are just covariant $z$ and $\bar{z}$ derivatives
respectively rescaled by $\sqrt{\theta}$.

Before looking for solutions to these equations it is interesting
to note that there are different symmetries of the equations of
motion.  The most obvious from \eom is the conjugation transformation
\eqn\symmone{D_{\mu} \rightarrow S D_{\mu} S^{\dagger},
\Phi_{a} \rightarrow S \Phi_{a} S^{\dagger}}
that we have been discussing so far.
In this setting we call this symmetry the ``non-BPS symmetry'' because
in general a solution to the BPS equations will not map back
to a solution of the BPS equations under this symmetry, but rather
only to a solution of the full equations of motion.
The BPS equations however have another
symmetry \refs{\hash, \ter} which takes 
one BPS solution to another BPS solution.
This follows by defining the field
\eqn\phip{\Phi^{(P)} := \Phi - {x^{3} \over \theta}.} 
The first BPS equation then becomes
\eqn\rewrittenBPS{
{\partial \Phi^{(P)} \over \partial x^{3}} 
= {1 \over \theta} [C,\bar{C}]}
while the remaining BPS equations take the same form with
$\Phi$ replaced by $\Phi^{(P)}$.  The BPS symmetry is then
given by conjugating all the fields as above, i.e.,
\eqn\BPSsymm{\Phi^{(P)} \rightarrow S \Phi^{(P)} S^{\dagger},
C \rightarrow S C S^{\dagger}}
and similarly for $\bar{C}$.

We now discuss a few simple solutions
to the BPS equations and the
solutions generated from them
under the BPS and non-BPS symmetry transformations respectively.
The simplest solution is the ground state given by
\eqn\vac{\Phi= \Phi_0 , C = a^{\dagger},
\bar{C} = a, B_{3}=0.} 
Acting on the ground state having  $ \Phi_0 = 0$ 
  with the BPS symmetry transformation 
one finds the fluxon solution \refs{\grnk,\hash,\ter}
\eqn\fluxon{\Phi = {x^{3} \over \theta} P_{0}, \, \, C=S a^{\dagger} 
S^{\dagger}, \, \, \bar{C} = S a S^{\dagger}, 
\, \, B_{3} = - {1 \over \theta}
P_{0}.}
 This solution corresponds to a D-string
piercing a D3-brane located in the $x^{1}$, $x^{2}$, $x^{3}$ plane.
The D-string is associated with a geometrical deformation 
 of the D3-brane worldvolume. The deformation is localized 
 near the origin of the $(x_1,x_2)$ plane, and takes the form 
 of a spike extending in the $ ( \Phi, x_3)$ plane 
 at an angle $ 1/\theta $  ( see for example \hash ).  
 Further evidence for this picture is given by considering 
 the charges of this configuration. 
The term $ \int C_{03} \wedge F_{12} $ shows that this 
 configuration has $D1$ charge in the $ 3$ direction. 
 The term $ \int C_{04} \wedge F_{12} \wedge [\partial_3, X^4 ] $
 coming from the non-abelian pull-back \myers\  
 gives D1-charge along the $\Phi \equiv X^4 $ direction. 
It is noteworthy that the latter charge has an extra factor of
 $ {1 \over \theta }  $ consistent, for large $\theta $ 
( where $ \tan(1/ \theta)  \sim { 1 \over \theta }$ ),  
 with the geometrical picture.  

Acting on \vac\
 with the non-BPS symmetry transformation one generates
the solution
\eqn\vacone{\Phi = \Phi_{0} (1-P_{0}), \, C=S a^{\dagger} S^{\dagger},
\, \bar{C} = S a S^{\dagger}, \,  B_{3} = {1 \over \theta} P_{0}}
This solution corresponds to a D3-brane located at $\Phi = \Phi_0$
along most of the $(x^{1},x^{2})$ plane and having a
deformation localized  near $x^{1}=0=x^{2}$ 
and extending to $ \Phi = 0 $. 

There are also interesting solutions 
 with $C=0=\bar{C}$ is given by
\eqn\Dstring{\Phi = \Phi_{0} + {x^{3} \over \theta}, C=0=\bar{C},
B_{3} = - {1 \over \theta}.}
This configuration carries infinite  D-string
charge along $x_3$ and along $\Phi$
through Chern-Simons couplings 
 discussed above.  This can be confirmed by a fluctuation 
 analysis which shows that momentum modes along the 
 $(x_1,x_2)$ directions have zero energy. 
We may also understand this picture 
 by noting that if one conjugates the ground
state solution \vac\ by  $S^{n}$ under the BPS symmetry then $\Phi$
becomes in that case $\Phi = (x^{3}/\theta) P_{n}$ where $P_{n}$
projects onto an $n$-dimensional subspace.  This solution is the 
$n$ fluxon solution corresponding to $n$ D-strings piercing the 
D3-brane.  As $n$ is taken to infinity one obtains \Dstring (with
$\Phi_{0}=0$).

Conjugating \Dstring\  by the BPS symmetry one finds
\eqn\DstringBPS{\Phi = (\Phi_{0} + {x^{3} \over \theta}) - \Phi_{0}
P_{0}, ~~
C=0=\bar{C}, ~~
B_{3} =- {1 \over \theta}}
while the non-BPS symmetry yields
\eqn\DstringnonBPS{\Phi = (\Phi_{0} + {x^{3} \over \theta})(1 -
P_{0}), ~~
C=0=\bar{C}, ~~
B_{3} =- {1 \over \theta}.}
The BPS solution represents 
a D3-brane with infinite D-string charge
inclined in the $(x_3, \Phi )$ plane 
 and shifted by $ \Phi_0$ everywhere 
 in the $x_1,x_2$ plane except near the origin. 
The non-BPS solution represents a D3-brane 
with infinite D-string charge inclined in the 
 $ x_3, \Phi $ plane and having a D-string 
 near the origin of the $x_1,x_2$ plane 
  intersecting it and pointing along $ \Phi = 0 $
 in the $ (\Phi, x_3 )$ plane.

\newsec{Interpolating from vacuum to fluxon }

%To embed the $U(1)$ vacuum \vac  in the $U(N)$ theory we can
%simply multiply the fields by the $N \times N$ identity matrix,
%i.e., the $U(N)$ vacuum solution is
%\eqn\unvac{\Phi = 0, \, D = -{a^{\dagger} \over \sqrt{2 \theta}}  
%I_{N \times N}, \, \bar{D} = {a \over \sqrt{2 \theta}} I_{N \times N}.}
%The BPS and non-BPS symmetries are generated in the same way
%as in the $U(1)$ case except that now the transformation
%operator $U$ is an $N \times N$ matrix with entries operators
%acting on the Fock space \Fock.

%As a specific example consider the $U(4)$ case with symmetry
%generator
%\eqn\Sfour{ S_{(4)} = diag(S,0,0,0).}

  Interpolating solutions can be constructed 
  after embedding in $U(4)$.  
 Consider a $U(4)$ 
 solution obtained by a diagonal embedding 
of one copy of \vacone\ and three copies of 
the solution in  \Dstring\ with $ \Phi_0 = 0 $.   
\eqn\sfoursol{\eqalign{  
\Phi &=  { x^3 \over \theta } Diag (P_{0}, 1, 1, 1) \cr 
\Phi^{(P)} & =  { - x^{3} \over \theta} diag (1-P_{0},0,0,0) \cr 
 B_3 &= { -1 \over \theta } Diag ( P_0, 1,1 , 1 ). 
  \cr }}
The first entry along the diagonal describes a fluxon. 
The remaining  entries describe sheets carrying infinite 
D-string charge at an angle in $ \Phi, x_3 $ plane. 

This is related to the following configuration 
 by BPS symmetry : 
\eqn\conto{\eqalign{  
\Phi &=  { x^3 \over \theta }  Diag ~~( 0,1,1,1) \cr 
\Phi^{(P)} &= - { x^3 \over \theta } Diag ~~ ( 1,0,0,0 ) \cr 
B_3 &= -{1\over \theta } Diag ~~ (0,1,1,1 ) \cr }} 
Multiplying \conto\ by $ W (t)$ of \homotopy\ 
 on the left and $ W^{-1}(t) $ on the right 
yields a family of solutions starting at $t=0$ from \conto\ 
and ending at $t=1$ at \sfoursol. 
The interpolation removes the D-string fluxon 
from the first three-brane, in the presence of 
three extra three-branes carrying infinite D-string charge.  

 One picture of what is
happening in the interpolation 
 between \sfoursol\ and \conto\
 is that the D-string breaks at the location
at which it intersects the D3-brane while the endpoints
remain attached to the D3-brane forming a monopole/anti- 
monopole pair.  Pulling the monopole(anti-monopole) to
$x^{3} = \infty (- \infty)$ respectively we arrive at the 
solution described above.  It would be interesting to
test this picture by explicit investigations
 of the interpolating configurations.

Interpolating families can be constructed 
 based on the non-BPS conjugation as well. 
 They connect, for example, 
\eqn\conbpi{\eqalign{
& \Phi = { x^3 \over \theta } Diag ~~( 1- P_0  ,0,0,0 )  \cr 
& B_3 = {1 \over \theta} Diag ~~( 1,1,1,1) \cr 
& C= \bar C = 0 \cr }}
to the solution 
\eqn\conbp{\eqalign{
& \Phi = { x^3 \over \theta } Diag ~~ ( 1,0,0,0 )  \cr 
& B_3 = {1 \over \theta} Diag ~~ ( 1,1,1,1) \cr 
& C= \bar C = 0 \cr }}
Both initial and final configurations 
 contain three 3-branes at $ \Phi = 0 $
 carrying infinite D-string charge. 
The initial configuration has in addition 
a tilted D3 with infinite D-string charge, 
with a D-string piercing it. 
The final configuration has the piercing D-string 
 removed.

\newsec{ Summary and Outlook } 

 We have looked at families of  static on-shell open string configurations 
 which interpolate between different classical solutions as well as
 time-dependent solutions of the Euclidean 
 equations of motion. Our interpolating technique 
 comes from a basic theorem in the K-theory 
 of operator algebras. It would be interesting 
 to find the 
 K-theoretic 
 significance ( along the lines of \refs{\wit,\witov,\minmo,\hamo,\ms }  )
  of the  fact that we can 
 interpolate between these non-commutative solutions
 when the rank of the gauge group becomes
 precisely such that the different definitions 
 of projector eqivalence become identical.  

We have so far not mentioned time-dependent solutions
for Minkowski signature.  In fact such solutions can
be generated in exactly the same manner as the
Euclidean solutions.  For example it is easy 
to generalize the interpolating solution
in \icfiii\ and \icfiiig\ to the case where $t$ is
the time coordinate in Minkowski space.  It turns out that 
if we perform a naive regulation of the trace by cutting off
 the states in the Hilbert space at finite level number to make
the energy well-defined, then this energy is not conserved.  
 This is trivial to see in that we start with the
mixed closed/open string vacuum configuration in 
\stconfii\ and end in the D23 brane, open string
vacua configuration in \ndconfii.  The existence
of formal solutions violating energy conservation 
can actually be seen in  the ordinary wave equation.  However
in that case these solutions are in general unphysical
and discarded.  In the case at hand
though we see
no obvious reason why such solutions should be
discarded.
We hope to explore the Minkowskian solutions further 
in the near future. 

 We do not expect the solutions discussed in 4.5 
 to be deformable to any  finite action ones but 
 it is conceivable that those in section 
 4.3 can be so deformed. It would be very interesting 
 to exhibit finite action solutions 
 which accomplish the kind of vacuum to brane
 or vacuum to vacuum via a brane that we described.  
 Repacing the NC $R^2$ by a NC $T^2$ would cure 
 some of the infinities we had but it seems unlikely 
 to cure all of them. 

  We made some comments on the relation 
  of these  families of non-commutative solitons 
  to the works of \hhk\senuniq\dgi\  in section 4. 
  We expect that further exploration of these families of NC solitons 
  will have interesting implications for the topology 
  of the configuration space of string fields. 
  While the existence of the unstable D-branes 
  can be understood by interpreting them 
  as sphalerons associated with certain instantons 
 in string theory, it is natural to ask, for example, 
   if there is a similar interpretation for 
 the existence of the families of interpolating solutions
constructed here.

\bigskip
 
 \noindent{\bf Acknowledgements:}
 We are specially grateful to Antal Jevicki for 
 many instructive discussions. 
 We are happy to acknowledge  useful  discussions with I. Brunner, 
 Z. Guralnik, D. Lowe. We have also benefitted from the lectures
 on K-theory for physicists by I. Singer at MIT \sing.  
 This research was supported by DOE grant  
 DE-FG02/91ER40688-(Task A).

\newsec{ Appendix }

 We review some facts about K-theory
 of operator algebras, highlighting 
 some concepts and formulae, while referring to standard 
 sources like \blacad\ for
 a detailed exposition of definitions 
 and proofs. These facts are used in the 
 construction of solutions in the text. 
 
 Let $A$ be an algebra of operators in a Hilbert space. 
 $K_0 (A) $, the K-group of an algebra  
 $A$ is defined in terms of equivalence 
 classes of projectors. 
 A projector in $A$ is an element $p \in A$ 
 which satisfies $ p^2 = p $. 
 It is the space of inequivalent projectors 
 in $ M_{\infty} ( A ) $, where $ M_{\infty} ( A )$ 
 is the algebra of large $N$ matrices with entries 
 taking values in $A$. There are different definitions
 of  equivalence. Two projectors $p$ and $q$ are said 
 to be algebraically equivalent if there are elements 
 $ x, y \in A $ such that $ xy = P, yx = Q, x = Px= xQ = PxQ,  
 y = Q y = y P = QyP $. Two projectors 
 $ P$ and $Q$ are said to be equivalent by similarity 
 if they conjugates by a unitary element $z$, 
 i.e $ z P z^{-1} = Q $. Two projectors are said to 
 be homotopically equivalent if there is a family of projectors
 connecting $P$ and $Q$. 

 While these definitions are not equivalent in $A$ 
 they become equivalent when we consider 
 matrices whose entries take values in $A$. 
 The first relevant result 
 is that if $P$ and $Q$ are algebraically equivalent, 
 then 
\eqn\alsim{ \pmatrix{ & P & 0 \cr 
                      & 0 & 0 \cr } 
      = Z  \pmatrix{ & Q & 0 \cr 
                      & 0 & 0 \cr  } Z^{-1} } 
where $ Z = \pmatrix { & y & 1 - Q \cr 
                       & 1 - P & x \cr } $
 and $ Z^{-1} = \pmatrix{ & x & 1 - e \cr 
                          & 1 - f & y \cr }$. 
 The second result is that if $ Z$ is invertible 
 then there is a path of invertibles 
 in $M_2( A ) $ from $1$ to $ Diag ( Z, Z^{-1} ) $. 
 The interpolating matrix $ W(t) $ is 
 $ Diag( x, 1 ). u(t). diag( y, 1 ). u(t)^{-1} $
 where $$ u(t) = \pmatrix{& Cos{ \pi t \over 2 } & 
- Sin { \pi t \over 2 } \cr 
 & Sin{ \pi t \over 2 } & Cos{ \pi t \over 2 } \cr }.$$

 These basic K-theoretic constructions 
 are used in section 3, where, in the simplest 
 cases, the role of $P,Q$ is played by 
 $ 1, 1 - P_0 $, and the role of 
 $ x,y$ is played by $ S, S^{\dagger}$.

\listrefs

\end